\documentclass[11pt]{article}

\usepackage[a4paper,margin=1in]{geometry}
\usepackage[T1]{fontenc}
\usepackage{amsmath,amssymb,amsfonts,amsthm}
\usepackage{txfonts}
\usepackage{graphicx,xcolor}
\usepackage[colorlinks=true,linkcolor=blue,citecolor=blue,urlcolor=blue]{hyperref}
\usepackage[numbers]{natbib}
\usepackage{braket}
\usepackage{epstopdf}
\usepackage{subfigure}
\usepackage{url}
\usepackage{array}
\usepackage{float}

\newtheorem{theorem}{Theorem}[section]

\newtheorem{lemma}[theorem]{Lemma}
\newtheorem{proposition}[theorem]{Proposition}

\theoremstyle{definition}
\newtheorem{definition}{Definition}

\begin{document}

\begin{center}
{\LARGE\bfseries Quantum Network of Assets (QNA): A Density-Operator Framework for Market Dependence and Structural Risk Diagnostics\par}
\vspace{1em}
Hui Gong$^{1}$, Akash Sedai$^{1}$, Francesca Medda$^{1}$\par
\vspace{0.5em}
$^{1}$Institute of Finance and Technology, University College London, London WC1E 6BT, UK\par
\vspace{0.5em}
\texttt{h.gong.12@ucl.ac.uk}, \texttt{akash.sharma@ucl.ac.uk}, \texttt{f.medda@ucl.ac.uk}\par
\vspace{0.5em}
March 31, 2026
\end{center}

\begin{abstract}
Classical correlation and rolling PCA summarize market dependence through covariance spectra, but they do not provide a unified operator representation for entropy, purity-based mixing, and standardized structural deviations built from rolling multi-feature trajectories. We propose the \emph{Quantum Network of Assets} (QNA), a quantum-inspired but non-physical density-operator framework in which normalized asset-level state vectors induce a time-varying market operator and an associated overlap network. The framework yields two structural diagnostics: the Entanglement Risk Index (ERI) and the Quantum Early-Warning Signal (QEWS). Using a stable NASDAQ--100 panel over 2020--2025, spanning the pandemic aftermath, the 2022 tightening cycle, and the 2025 tariff repricing episode, we show that QNA entropy remains strongly related to covariance spectral entropy and effective rank at the regime level, but that the method becomes empirically distinct once the operator is constructed from multi-feature rolling trajectories rather than returns alone. In the returns-only limit, QNA lies close to classical spectral summaries; with volatility and liquidity channels included, it captures broader dependence reconfiguration and produces the clearest incremental signal during the April~2025 tariff escalation, when QEWS shifts sharply while rolling-z classical spectral benchmarks move only modestly. QNA therefore does not replace covariance-spectrum methods; instead, it provides a unified operator representation in which entropy, purity-based mixing, and event-aligned structural deviations are analyzed jointly across rolling multi-feature market states.
\end{abstract}



\section{Introduction}

The covariance matrix and its derived correlation coefficients form the backbone of modern portfolio theory, risk management, and financial network analysis. 
Despite their widespread use, classical correlation measures capture only \textit{linear} and \textit{pairwise} co-movements, implicitly assuming separability and stability of dependencies across market states. 
A broad empirical literature on correlation hierarchies, random-matrix filtering, connectedness, and eigenvalue-based concentration diagnostics shows that financial markets exhibit non-linear amplification, volatility clustering, and abrupt regime transitions, reflecting the behaviour of complex, interacting systems rather than collections of weakly coupled assets \cite{mantegna_hierarchical,mantegna_stanley,bonanno_equities,tumminello_networks,laloux_noise,plerou_random,billio_connectedness,diebold_network,barunik_krehlik,kritzman_systemic}. 
During periods of uncertainty or systemic stress, dependencies tend to concentrate, and the effective dimensionality of market behaviour collapses in ways that classical correlation matrices cannot adequately represent.

These limitations motivate the development of alternative representations of dependence that move beyond incremental refinements of covariance-based models. 
In this paper, we adopt an information-theoretic and network-oriented perspective, introducing a framework that generalises classical correlation by capturing higher-order, non-linear, and state-dependent interactions. 
Rather than modelling dependencies solely through second moments, the proposed approach characterises market structure through the geometry of information sharing across assets.

We introduce the \textit{Quantum Network of Assets} (QNA), an information-theoretic representation in which rolling asset-level feature trajectories are embedded into a density-matrix formalism. 
This construction is \emph{quantum-inspired but not physical}: no assumptions are made about quantum behaviour in financial markets. 
Instead, the framework uses the mathematical properties of density operators, entropy, and mutual information to describe cross-asset dependence in a normalized operator representation. 
Mathematically, the resulting object is closely related to a Gram/covariance operator built from standardized rolling feature blocks, so its eigenspectrum is naturally connected to rolling PCA, effective-rank, absorption-ratio, and spectral-entropy summaries \cite{jolliffe_cadima,litterman_scheinkman,roy_vetterli,kritzman_systemic}. 
We therefore treat classical spectral methods as benchmarks rather than foils: QNA is intended to complement them by providing a common representation for entropy, purity, standardized structural deviations, and partition-based extensions.

Within the QNA framework, we derive two structural indicators that summarise different aspects of market organisation:

\begin{itemize}
    \item \textbf{Entanglement Risk Index (ERI)}, a scalar purity-based summary of structural mixing and non-separable cross-asset dependence. 
    We retain the acronym for continuity, but interpret it as an operator-based dependence measure rather than as a claim about physical entanglement.

    \item \textbf{Quantum Early-Warning Signal (QEWS)}, defined from the temporal dynamics of entropy-based structural measures. 
    QEWS is designed as a diagnostic indicator of abnormal structural deviation relative to recent history, rather than as a predictive trading signal.
\end{itemize}

The emphasis of this study is therefore \textit{structural rather than predictive}. 
The proposed indicators are intended to characterise how dependence geometry evolves through time, not to forecast prices or returns. 
In particular, the framework highlights changes in effective dimensionality, structural mixing, and abnormal departures from recent operator states. 
These phenomena are consistent with the view of financial markets as adaptive systems in which dependence reconfiguration often accompanies, and sometimes leads, observable price adjustment.

Using empirical data for a stable NASDAQ--100 panel over 2020--2025, we benchmark QNA against covariance spectral entropy, effective rank, and participation-ratio summaries \cite{jolliffe_cadima,roy_vetterli,kritzman_systemic}. 
The revised evidence shows strong regime-level comovement between QNA entropy and classical spectral measures, but only moderate alignment between QEWS and rolling-z classical benchmarks. 
This distinction is central rather than incidental: returns-only QNA behaves much like a normalized spectral covariance diagnostic, whereas the multi-feature specification departs from that limit and supports a broader operator interpretation of dependence reconfiguration. 
QNA therefore does not replace covariance-spectrum methods; instead, it provides a unified operator representation in which entropy, purity-based mixing, and event-aligned structural deviations are analyzed jointly across rolling multi-feature market states. 
Empirically, we study the full 2020--2025 sample through three spotlight regimes -- the 2020 pandemic aftermath, the 2022 macro-tightening repricing, and the 2025 tariff cluster -- while visually emphasizing 2024--2025 because it provides the cleanest non-crisis setting in which structural reconfiguration can be separated from broad volatility spikes. We also report explicit sensitivity analyses across feature sets and 40/60/90-day rolling windows, using 40 days as the shortest horizon that preserves stable multi-feature coverage, and we replace the earlier single-event discussion with a broader event-window design spanning Fed liftoff, Jackson Hole, a two-stage 2025 tariff episode, and placebo windows.

The remainder of the paper is organised as follows. 
Section~2 introduces the theoretical foundations of the information-theoretic framework. 
Section~3 formalises the Quantum Network of Assets and derives the associated structural indicators. 
Section~4 describes the data and implementation, and Section~5 presents the empirical results. 
Section~6 discusses implications for systemic risk and financial stability, and Section~7 concludes. 
Overall, QNA does not replace classical correlation analysis; rather, it provides a principled generalisation that complements existing tools by revealing hidden structure in financial dependency networks.

\section{Theoretical Foundations}

\subsection{From Classical Correlation to Information-Theoretic States}

Classical correlation matrices quantify linear co-movement between pairs of assets. 
Given demeaned returns $r_i(t)$, the correlation between assets $i$ and $j$ is defined as
\begin{equation}
\rho_{ij} =
\frac{\mathbb{E}[r_i r_j]}
{\sqrt{\mathbb{E}[r_i^2] \; \mathbb{E}[r_j^2]}}.
\label{eq:classical_corr}
\end{equation}
This formulation relies on three implicit assumptions:
\begin{enumerate}
    \item dependence is linear,
    \item interactions are pairwise and separable,
    \item the dependency structure evolves slowly over time.
\end{enumerate}
Empirical evidence suggests that financial markets frequently violate all three, particularly during periods of stress, policy uncertainty, or regime transition, when dependencies become non-linear, clustered, and state-dependent.

\vspace{0.5em}
\noindent\textbf{Embedding returns into an information-theoretic state representation.}

To move beyond pairwise correlation, the QNA framework embeds cross-asset information into a state-based representation inspired by quantum information theory. 
Given a normalised return vector $\psi(t)$ with components
\begin{equation}
\psi_i(t) = 
\frac{r_i(t)}{\sqrt{\sum_{j=1}^N r_j^2(t)}},
\label{eq:psi_normalisation}
\end{equation}
we define a state vector
\begin{equation}
\ket{\psi(t)} = \sum_{i=1}^N \psi_i(t) \ket{i},
\label{eq:psi_superposition}
\end{equation}
where the basis $\{\ket{i}\}$ labels assets rather than physical states.

The associated density operator is
\begin{equation}
\rho(t) = \ket{\psi(t)}\bra{\psi(t)},
\label{eq:density_matrix}
\end{equation}
with elements
\begin{equation}
\rho_{ij}(t) = \psi_i(t)\psi_j(t).
\label{eq:density_components}
\end{equation}

The diagonal elements $\rho_{ii}$ represent relative concentration across coordinates of the chosen state representation, while the off-diagonal elements encode joint structure arising from the full configuration of normalized asset trajectories. 
These terms do not map one-for-one into contemporaneous pairwise covariance coefficients.

\vspace{0.5em}
\noindent\textbf{Why density matrices generalise covariance matrices.}

A classical covariance matrix,
\begin{equation}
\Sigma_{ij} = \mathbb{E}[r_i r_j],
\label{eq:covariance}
\end{equation}
captures only second-moment dispersion.

In contrast, the density matrix $\rho$ provides access to:
\begin{itemize}
    \item global amplitude-based dependence,
    \item non-separable cross-asset structure,
    \item entropy-based measures of effective dimensionality.
\end{itemize}

Following the information-theoretic framework of \cite{shannon_communication,nielsen_chuang}, this representation enables the computation of:
\begin{itemize}
    \item von Neumann entropy $S(\rho)$,
    \item mutual information between asset partitions,
    \item structural mixing and concentration.
\end{itemize}

These quantities characterise collective behaviour—such as dimensionality compression and information synchronisation—that remain hidden under covariance-based analysis.

\subsection{Non-Separable Dependence in Financial Networks}

The QNA framework does not posit physical quantum entanglement in financial markets. 
Instead, it adopts the mathematical notion of \emph{non-separability} as a precise descriptor of structural dependence. 
Once the market is represented by a density matrix $\rho$, interactions among asset groups can be analysed at the subsystem level.

\vspace{0.5em}
\noindent\textbf{Subsystems and market partitions.}

Let the market be partitioned into two subsets $A$ and $B$ (e.g.\ sectors, regions, or size groups). 
From the joint state $\rho_{AB}$, we define reduced states
\begin{equation}
\rho_A = \mathrm{Tr}_B(\rho_{AB}), 
\qquad
\rho_B = \mathrm{Tr}_A(\rho_{AB}),
\label{eq:partial_trace}
\end{equation}
where $\mathrm{Tr}$ denotes partial tracing.

Even when the global state is pure, the reduced states are typically mixed, reflecting statistical dependence between subsystems. 
This property holds regardless of whether dependencies are linear or non-linear.

\vspace{0.5em}
\noindent\textbf{Entropy and effective degrees of freedom.}

The von Neumann entropy,
\begin{equation}
S(\rho) = - \mathrm{Tr}(\rho \log \rho),
\label{eq:theory_von_neumann_entropy}
\end{equation}
measures the effective dimensionality of the market state.

In financial terms:
\begin{itemize}
    \item high entropy corresponds to diversified, multi-factor structures,
    \item low entropy reflects concentration and synchronisation of risk channels.
\end{itemize}

Falling entropy therefore signals structural concentration or tightening, while rising entropy reflects a more mixed multi-mode market state.

\vspace{0.5em}
\noindent\textbf{Mutual information and structural coupling.}

The mutual information between partitions $A$ and $B$ is defined as
\begin{equation}
I(A\!:\!B) = S(\rho_A) + S(\rho_B) - S(\rho_{AB}).
\label{eq:quantum_mi}
\end{equation}
Unlike classical correlation, this measure captures total dependence, including non-linear and higher-order interactions.

Large values of $I(A\!:\!B)$ indicate strong structural coupling, meaning that the behaviour of one subset cannot be described independently of the other. 
This notion of non-separability forms the basis of the Entanglement Risk Index (ERI).

\subsection{Scope of the Analogy}

The non-separable dependence encoded in $\rho$ admits a formal analogy with quantum-information concepts, but the role of that analogy in this paper is strictly mathematical. 
The manuscript does not claim Bell-type effects or physical entanglement in financial markets. 
Instead, the analogy serves to distinguish separable from non-separable dependency structures in a compact operator language.

In classical probability, separability implies factorisation:
\begin{equation}
P(a,b) = P(a)P(b).
\label{eq:classical_factorization}
\end{equation}
By contrast, a joint density matrix may satisfy
\begin{equation}
\rho_{AB} \neq \sum_k p_k \, \rho_A^{(k)} \otimes \rho_B^{(k)},
\label{eq:nonseparable}
\end{equation}
indicating structural dependence beyond classical factor models.

The relevance lies in the shared mathematical structure: global operator states can encode dependence patterns that are not reducible to independent components or to a small set of contemporaneous pairwise summaries.

\vspace{0.5em}
\noindent\textbf{Interpretation for financial markets.}

Non-separable dependence corresponds to:
\begin{enumerate}
    \item synchronisation of expectations across assets,
    \item compression of effective market dimensionality,
    \item heightened sensitivity to common information shocks.
\end{enumerate}

Major announcements can then be interpreted as points at which latent dependence structure is reconfigured. 
QEWS tracks these transitions by monitoring standardized departures of entropy or ERI from recent history.

\subsection{Latent Information and Structural Transitions}

Prior to public disclosure, markets often reflect overlapping and partially formed expectations. 
Within QNA, this corresponds to a mixed density matrix,
\begin{equation}
\rho = \sum_k p_k \ket{\psi_k}\bra{\psi_k},
\label{eq:mixed_state}
\end{equation}
representing competing structural configurations.

As information accumulates, the state operator can become either more concentrated or more dispersed depending on how market modes reorganize. 
For that reason, the sign of QEWS is not interpreted mechanically as ``risk up'' or ``risk down''; rather, its magnitude and timing indicate abnormal structural adjustment relative to recent history. QEWS should therefore be interpreted as a signed deviation from a recent dependence baseline, not as a monotonic measure of systemic stress.
Following disclosure, entropy and ERI may normalize or reconfigure sharply as uncertainty resolves. 
QEWS captures this pre- and post-event adjustment without invoking predictive modelling.

\paragraph{Summary}
Section~2 establishes QNA as an information-theoretic extension of classical dependence analysis. 
By moving from pairwise covariance summaries to a state-based operator representation, the framework provides access to structural diagnostics---entropy, mutual information, purity, and QEWS---that can be compared directly with spectral baselines while retaining a richer language for regime reconfiguration.

\section{The Quantum Network of Assets (QNA)}

\subsection{Market State as a Density Matrix}

In the QNA framework, each asset is mapped to a normalised feature (return) vector
over a chosen window, and the cross-sectional market configuration is represented
by a density matrix. This construction is \emph{mathematical} rather than physical:
it uses the operator language of quantum information as an information-theoretic
device for encoding high-dimensional dependence and for enabling entropy-based
structural diagnostics.

\begin{definition}[Rolling Feature Amplitude Vector]
Let $x_i(t)\in\mathbb{R}^{dT}$ denote the concatenated, window-standardised
feature trajectory of asset $i$ over a rolling window of length $T$ with $d$
feature channels. In the baseline specification, the channels are returns,
volatility proxy, log volume, and log-volume change, stacked over the full
window. Define the normalised amplitude vector
\begin{equation}
    \ket{\psi_i}
    = \frac{x_i(t)}{\|x_i(t)\|_2},
\label{eq:amplitude_def}
\end{equation}
so that $\braket{\psi_i|\psi_i}=1$.
\end{definition}

This normalisation places heterogeneous assets on a common scale and can be
viewed as embedding rolling-window feature behaviour onto the unit sphere.
The notation follows the standard density-operator formalism used in information
theory (e.g.\ \cite{shannon_communication,nielsen_chuang}), but the basis vectors here index time-feature
coordinates rather than physical states.

\begin{definition}[Market Density Matrix]
Given $N$ assets with amplitude vectors $\{\ket{\psi_i}\}_{i=1}^N$,
define the \emph{market density matrix} by
\begin{equation}
    \rho
    = \frac{1}{N}\sum_{i=1}^N \ket{\psi_i}\bra{\psi_i}.
\label{eq:density_matrix_def}
\end{equation}
\end{definition}

By construction, $\rho$ is Hermitian, positive semidefinite, and satisfies
$\mathrm{Tr}(\rho)=1$. It therefore defines a valid density operator describing
an empirical ensemble of asset states.

\begin{lemma}[Cross-Asset Structural Coupling]
Let $\rho$ be defined by \eqref{eq:density_matrix_def}. Its entries satisfy
\begin{equation}
    \rho_{jk}
    = \frac{1}{N}\sum_{i=1}^N \psi_i(j)\,\psi_i(k),
\label{eq:coherence_terms}
\end{equation}
where off-diagonal terms $(j\neq k)$ encode cross-sectional structural coupling
that is not captured by covariance alone.
\end{lemma}

\begin{proof}
Expanding $\ket{\psi_i}\bra{\psi_i}$ yields
$\left[\ket{\psi_i}\bra{\psi_i}\right]_{jk}=\psi_i(j)\psi_i(k)$.
Averaging over $i$ gives \eqref{eq:coherence_terms}.
\end{proof}

\noindent
\textbf{Interpretation.}
The operator $\rho$ aggregates the cross-sectional structure of the market into a
single object. Its eigen-spectrum summarises:
\begin{itemize}
    \item the \emph{effective dimensionality} of market behaviour (number of active modes),
    \item the degree of \emph{structural concentration} versus dispersion of risk channels,
    \item the extent of \emph{non-separable dependence} across assets.
\end{itemize}
These properties are difficult to recover reliably from pairwise correlations,
especially when dependence is state-dependent or non-linear.

\vspace{0.5em}
\noindent
\textbf{Why the construction is quantum-inspired and why it is a network.}
The ``quantum'' content of QNA lies in the mathematical architecture rather than
in any claim about physical quantum behavior. Asset trajectories are embedded as
unit-norm state vectors, dependence is summarized by a density operator, and the
same operator supports entropy, purity, partial-trace, and mutual-information
calculations within a single Hilbert-space-like representation. The ``network''
content arises because the normalized amplitudes induce a time-varying overlap
kernel
\begin{equation}
    G_{ij}(t)=\braket{\psi_i(t)|\psi_j(t)},
\label{eq:overlap_kernel}
\end{equation}
which defines a fully weighted dependence network in asset space. The density
operator $\rho(t)$ is the state-space compression of that network geometry: it
retains the common dependence modes of the overlap structure while remaining
compatible with operator-based information measures.

\vspace{0.5em}
\noindent
\textbf{Relation to PCA and covariance-spectrum methods.}
If $\Psi(t)$ stacks the unit-norm asset vectors row-wise, then the market-state
operator can be written as $\rho(t)=\Psi(t)^{\top}\Psi(t)/N$. Its non-zero
eigenvalues coincide with those of the asset-space Gram matrix
$G(t)=\Psi(t)\Psi(t)^{\top}/N$, so rolling PCA, effective-rank measures, and
spectral-entropy diagnostics are natural benchmarks. The difference is not that
QNA escapes spectral analysis, but that it applies spectral analysis to a
normalized rolling-feature representation and then uses the same operator to
construct purity-based and partition-based diagnostics. In the returns-only
limit, this representation approaches the classical covariance-spectrum
benchmark; once additional channels are introduced, the operator can track
dependence reconfiguration that is not reducible to contemporaneous return
covariance alone.

\begin{proposition}[Separable Limit and the Classical Baseline]
If the asset amplitude vectors $\{\ket{\psi_i}\}$ are mutually orthogonal, then
$\rho$ is diagonal in that basis and entropy-based measures reduce to their
classical (Shannon-type) counterparts. This corresponds to a limiting case in
which cross-sectional coupling vanishes.
\end{proposition}

\begin{proof}
Orthogonality implies $\braket{\psi_i|\psi_j}=0$ for $i\neq j$. Hence
\eqref{eq:density_matrix_def} becomes diagonal in the basis $\{\ket{\psi_i}\}$,
and $S(\rho)$ reduces to the Shannon entropy of the eigenvalue distribution.
\end{proof}

\subsection{Entropy and Mutual Information}

Entropy and mutual information provide compact summaries of the market's
structural configuration encoded in $\rho$. Unlike classical correlation, which
targets second-order comovement, these information-theoretic quantities are
defined directly from the spectrum of $\rho$ and remain meaningful under
non-linear or unstable dependence.

\begin{definition}[Von Neumann Entropy]
For a market density matrix $\rho$, define
\begin{equation}
    S(\rho)
    = -\mathrm{Tr}(\rho\log\rho)
    = -\sum_{k=1}^T \lambda_k \log \lambda_k,
\label{eq:von_neumann_entropy}
\end{equation}
where $\{\lambda_k\}$ are the eigenvalues of $\rho$.
\end{definition}

In the market context, $S(\rho)$ can be interpreted as an \emph{effective
dimensionality} measure:
high entropy reflects dispersed, multi-mode structure; low entropy corresponds to
concentration into a small number of dominant modes (i.e.\ synchronisation).

\begin{definition}[Mutual Information Between Market Partitions]
Let the asset universe be partitioned into two groups $A$ and $B$ and let
$\rho_A,\rho_B$ denote reduced density matrices. Define
\begin{equation}
    I(A\!:\!B)
    = S(\rho_A) + S(\rho_B) - S(\rho_{AB}).
\label{eq:qmi_definition}
\end{equation}
\end{definition}

$I(A\!:\!B)\ge 0$ measures total dependence between partitions (including
non-linear and higher-order components) without requiring Gaussianity or linear
factor structure. In QNA, it quantifies structural coupling and information
synchronisation across groups of assets.

\vspace{0.5em}
\noindent
\textbf{Relation to classical baselines.}
When $\rho$ is diagonal (no cross-sectional coupling), \eqref{eq:von_neumann_entropy}
reduces to Shannon entropy and \eqref{eq:qmi_definition} reduces to classical mutual
information. Hence, covariance-based analysis appears as a separable limiting case.

\subsection{Purity Index and Entanglement Risk Index (ERI)}

To summarise the overall degree of structural mixing in the market state, QNA
uses the \emph{purity} of $\rho$. We retain the term ``Entanglement Risk Index''
for continuity, but interpret it strictly as a scalar measure of
\emph{non-separable dependence} (structural mixing), not physical entanglement.

\begin{definition}[Purity (``Quantum Index'')]
Define the purity of the market state as
\begin{equation}
    Q(t) = \mathrm{Tr}\bigl(\rho(t)^2\bigr).
\label{eq:quantum_index}
\end{equation}
\end{definition}

Purity satisfies $\frac{1}{dT}\le Q(t)\le 1$. Higher purity indicates greater
concentration into a smaller number of dominant modes, while lower purity
indicates a more mixed operator spectrum.

\begin{definition}[Entanglement Risk Index (ERI)]
Define
\begin{equation}
    \mathrm{ERI}(t) = 1-\mathrm{Tr}\bigl(\rho(t)^2\bigr) = 1-Q(t).
\label{eq:eri_definition}
\end{equation}
\end{definition}

High ERI corresponds to stronger structural mixing and a more dispersed operator
spectrum, whereas low ERI corresponds to greater concentration into a smaller
number of common modes. Interpreted financially, low-ERI states are more
consistent with:
\begin{itemize}
    \item tighter cross-asset coupling,
    \item erosion of diversification capacity,
    \item compression of effective market degrees of freedom.
\end{itemize}

\vspace{0.5em}
\noindent
\textbf{Relation to entropy.}
Both $S(\rho)$ and ERI reflect how mass is distributed across the operator
spectrum:
entropy summarises the full eigenspectrum, while ERI (via purity) places greater
weight on concentration in dominant modes. For empirical interpretation, it is
therefore useful to read entropy, purity, and QEWS jointly rather than forcing
any single monotone ``stress'' ordering on one index alone.

\subsection{Quantum Early-Warning Signal (QEWS)}

To detect \emph{structural transitions} rather than to forecast prices, we
define a standardised deviation of entropy (or ERI) relative to recent history.

\begin{definition}[Quantum Early-Warning Signal]
Let $S(t)$ denote entropy and $\mathrm{ERI}(t)$ denote the mixing index at time $t$.
Define
\begin{equation}
    \mathrm{QEWS}(t)
    = \frac{S(t) - \mu_S(t)}{\sigma_S(t)},
\label{eq:qews_entropy}
\end{equation}
or equivalently
\begin{equation}
    \mathrm{QEWS}_{\mathrm{ERI}}(t)
    = \frac{\mathrm{ERI}(t) - \mu_{\mathrm{ERI}}(t)}
           {\sigma_{\mathrm{ERI}}(t)},
\label{eq:qews_eri}
\end{equation}
where $\mu(\cdot)$ and $\sigma(\cdot)$ are rolling estimates over a window of
length $w$.
\end{definition}

QEWS highlights \emph{unusual} deviations in dependency geometry relative to
recent conditions. Positive values indicate above-baseline entropy or structural
mixing, while negative values indicate below-baseline values (greater
concentration). Importantly, QEWS is not designed as a predictive trading
signal; it is a diagnostic indicator of latent structural change.

\paragraph{Interpretation}
In practice, large-magnitude QEWS indicates that the market is entering a
configuration that differs materially from its recent dependence baseline.
The sign records direction; the magnitude records abnormality.

\section{Data and Implementation}

\subsection{Data}

We study the NASDAQ--100 equity universe using daily observations from
January~2020 to December~2025. We choose this sample because it is the shortest
window that simultaneously captures three qualitatively distinct dependence
regimes that matter for identification: the pandemic-era collapse in effective
dimensionality, the 2022 policy-tightening repricing, and the 2025 tariff-driven
dependence reconfiguration. In that sense, 2020--2025 is the \emph{main sample},
not a convenience window. The manuscript nevertheless gives special visual
attention to 2024--2025 because that segment provides a comparatively
non-crisis environment in which structural reconfiguration can be separated more
cleanly from the broad volatility shock that dominates early 2020. The pandemic
period remains visible in the full-sample benchmark figure, while 2022 and 2025
enter the formal event-study design. The empirical objective is \emph{structural
measurement} rather than forecasting: we reconstruct a time-varying market
state $\rho(t)$ and compute information-theoretic indicators that summarise
cross-sectional dependence and regime changes.

The ticker roster is anchored to the current NASDAQ--100 membership available at
download time, so the empirical panel should be interpreted as a stable large-cap
technology/growth universe rather than as a reconstruction of historical index
membership. This design avoids composition breaks inside the rolling operator,
but it also introduces survivorship bias; we therefore treat the results as
diagnostics on a stable panel and revisit this limitation in Section~6.

Daily adjusted close prices and trading volumes are retrieved via \texttt{yfinance}.
All series are aligned to a common U.S.\ business-day calendar. Days with missing
observations for a given asset are handled conservatively: within each rolling
window we require a minimum coverage threshold (e.g.\ at least $90\%$ of trading
days); assets failing this requirement are excluded for that window to avoid
introducing spurious dependence through aggressive imputation.\footnote{In the
baseline specification, isolated missing values are forward-filled for
\textit{prices} prior to return computation; volume series are not forward-filled
across extended gaps. Results are qualitatively robust to stricter filtering.}

\subsection{Rolling-Window Design}

All quantities are computed on a baseline rolling window of length $T=60$
trading days.
Let $\mathcal{U}(t)$ denote the set of assets with valid data in the window
$\{t-T+1,\ldots,t\}$ and let $N(t)=|\mathcal{U}(t)|$.
The rolling design serves two purposes: (i) it stabilises estimation of the
state operator $\rho(t)$ and its eigen-spectrum, and (ii) it enables the
structural indicators to evolve smoothly while remaining responsive to regime
changes. We use 60 days as the baseline compromise between responsiveness and
operator stability in the multi-feature specification; 40-day windows are the
shortest robustness horizon retained in the paper because they preserve more
stable feature-block coverage than a 30-day alternative, while 90-day windows
emphasize persistence at the cost of slower event response.

\subsection{Pipeline Overview}

Figure~\ref{fig:qna_pipeline} summarises the computational workflow. Each stage
is deterministic and fully reproducible given the input time series. The figure
also makes clear that QNA is a transformation of rolling feature blocks into an
operator representation rather than a parametric forecasting model.

\begin{figure}[h!]
\centering
\[
\begin{array}{c}
\text{Daily Prices \& Volumes} \\
\downarrow \\
\text{Returns \& Feature Standardisation} \\
\downarrow \\
\text{Windowed Feature Vectors } x_i(t) \\
\downarrow \\
\text{Normalised Amplitudes } \ket{\psi_i(t)} \\
\downarrow \\
\text{State Operator } \rho(t) \\
\downarrow \\
\{S(\rho(t)),\ I(A\!:\!B),\ \mathrm{ERI}(t),\ \mathrm{QEWS}(t)\}
\end{array}
\]
\caption{QNA processing pipeline: from market data to structural dependence indicators.}
\label{fig:qna_pipeline}
\end{figure}

\subsubsection*{Step 1: Returns and Basic Quantities}

For each asset $i \in \mathcal{U}(t)$ we compute log returns
\[
r_i(\tau)=\log P_i(\tau)-\log P_i(\tau-1),
\]
where $P_i(\tau)$ is the adjusted close price. We also retain daily trading volume
$V_i(\tau)$ and construct a windowed volatility proxy using a rolling standard
deviation of returns (computed within the same $T$-day window).

\subsubsection*{Step 2: Feature Construction and Standardisation}

At each time $t$, each asset is represented by a rolling feature block
\begin{equation}
x_i(t)=
\bigl(
    \tilde r_i(t-T+1{:}t),\;
    \tilde \sigma_i(t-T+1{:}t),\;
    \tilde v_i(t-T+1{:}t),\;
    \widetilde{\Delta v}_i(t-T+1{:}t)
\bigr),
\label{eq:feature_vector}
\end{equation}
where each block denotes a length-$T$ trajectory that has been standardised
within the rolling window. Thus $\tilde r_i(t-T+1{:}t)$ denotes the
window-standardised return path, $\tilde \sigma_i(t-T+1{:}t)$ the
window-standardised volatility path, $\tilde v_i(t-T+1{:}t)$ the
window-standardised log-volume path, and
$\widetilde{\Delta v}_i(t-T+1{:}t)$ a window-standardised volume-change (or
``volume acceleration'') path.\footnote{In the baseline implementation,
$\Delta v_i(t)=\log V_i(t)-\log V_i(t-1)$ and each component is standardised within
the rolling window to zero mean and unit variance to avoid scale dominance.}
This feature set is intentionally parsimonious: it introduces liquidity/activity
information without turning the construction into a large latent-factor model.
Because the operator depends on feature design by construction, the revision
reports explicit sensitivity checks for returns-only, reduced-feature, and
baseline multi-feature specifications rather than relegating them to
``available on request'' status.

\subsubsection*{Step 3: Amplitude (Unit-Norm) Representation}

To construct comparable state vectors across assets, we map each feature vector to
the unit sphere:
\begin{equation}
\ket{\psi_i(t)} = \frac{x_i(t)}{\|x_i(t)\|_2}.
\label{eq:amplitude_construct}
\end{equation}
This normalisation removes scale effects and ensures that differences across assets
reflect \emph{directional} configuration in feature space rather than magnitude.
In the operator formalism, $\ket{\psi_i(t)}$ plays the role of an empirical
``amplitude'' encoding each asset's local configuration at time $t$.

\subsubsection*{Step 4: State Operator (Density Matrix) Estimation}

Given $\{\ket{\psi_i(t)}\}_{i\in\mathcal{U}(t)}$, the market state is estimated by
\begin{equation}
\rho(t)=\frac{1}{N(t)}\sum_{i\in\mathcal{U}(t)} \ket{\psi_i(t)}\bra{\psi_i(t)}.
\label{eq:rho_estimation}
\end{equation}
This estimator is a sample average of rank-one projectors and is therefore
positive semidefinite with $\mathrm{Tr}(\rho(t))=1$. Its spectrum provides a
compact summary of the cross-sectional configuration of the market at time $t$.
Equivalently, the overlap matrix $G_{ij}(t)=\braket{\psi_i(t)|\psi_j(t)}$
defines the time-varying weighted asset network, while $\rho(t)$ compresses that
network into a state operator on which entropy, purity, and partition-based
diagnostics can be computed consistently.

\subsubsection*{Step 5: Entropy and Mutual Information}

We compute the von Neumann entropy
\[
S(\rho(t))=-\mathrm{Tr}\bigl(\rho(t)\log\rho(t)\bigr)
          =-\sum_k \lambda_k(t)\log\lambda_k(t),
\]
where $\{\lambda_k(t)\}$ are eigenvalues of $\rho(t)$.
To quantify dependence between market partitions, we compute mutual information
$I(A\!:\!B)$ using reduced density matrices obtained by tracing out the complement
partition (as defined in Section~2). In the present revision, partition-based
results are treated as supplementary diagnostics rather than as the headline
empirical contribution; the main text focuses on entropy, purity/ERI, QEWS, and
their benchmark comparison with classical spectral measures.

\subsubsection*{Step 6: Structural Mixing Index (ERI)}

We compute purity $Q(t)=\mathrm{Tr}(\rho(t)^2)$ and define
\[
\mathrm{ERI}(t)=1-Q(t)=1-\mathrm{Tr}\bigl(\rho(t)^2\bigr),
\]
which summarises the degree of structural mixing (non-separable dependence) in
the market state.

\subsubsection*{Step 7: Standardised Structural Deviation (QEWS)}

To detect abnormal structural transitions relative to recent history, we compute
the standardised deviation of ERI (or entropy) over a rolling window of length $w$:
\[
\mathrm{QEWS}(t)=
\frac{\mathrm{ERI}(t)-\mu_{\mathrm{ERI}}(t)}{\sigma_{\mathrm{ERI}}(t)}.
\]
QEWS is interpreted as a diagnostic indicator of structural change rather than as
a predictive model.

\subsection{Implementation and Numerical Details}

All computations are implemented in Python using \texttt{numpy} and \texttt{pandas},
with eigen-decompositions computed via standard linear algebra routines.
To ensure numerical stability in entropy calculations, eigenvalues below a small
tolerance $\epsilon$ are clipped (e.g.\ $\lambda \leftarrow \max(\lambda,\epsilon)$)
before applying $\log(\cdot)$.\footnote{This is a standard numerical safeguard that
does not affect results materially because the clipped mass is negligible.}

The codebase is modular: data ingestion, feature construction, state estimation,
and metric computation are implemented as separate functions. This design
facilitates extensions to other equity universes, alternative feature sets,
different window lengths, or higher-frequency data.

\section{Empirical Results}

This section addresses the three empirical questions raised most directly by the
review process. First, how closely does QNA track familiar covariance-spectrum
diagnostics? Second, how much of the apparent novelty depends on feature-block
design and rolling-window length? Third, do QEWS excursions remain visible once
the original descriptive tariff figure is replaced by a broader event-window
design with formal inference and placebo dates?

\subsection{Benchmark Relationships and Main-Sample Dynamics}

The first result is that QNA should indeed be read as a \emph{spectrally related}
representation rather than as a wholesale rejection of PCA-style methods
\cite{jolliffe_cadima,roy_vetterli,kritzman_systemic}.
Table~\ref{tab:benchmark_summary} reports full-sample correlations between the
QNA diagnostics and classical dependence summaries. At the level of the state
operator, QNA entropy is strongly related to covariance spectral entropy
($0.823$), effective rank ($0.801$), and participation ratio ($0.737$); ERI
shows the same qualitative pattern. These are not small residual associations:
they confirm that the QNA eigenspectrum shares substantial regime information
with rolling covariance-spectrum methods.

\begin{table}[h!]
\centering
\small
\caption{Benchmark correlations between QNA diagnostics and classical dependence summaries. Entries report full-sample correlations with bootstrap 95\% confidence intervals.}
\label{tab:benchmark_summary}
\resizebox{\textwidth}{!}{%
\begin{tabular}{lccccc}
\hline
\textbf{Panel A: Level diagnostics} & \textbf{Cov. spectral entropy} & \textbf{Effective rank} & \textbf{Participation ratio} & \textbf{Mean abs. corr.} & \textbf{Realized vol.} \\
\hline
        ERI & 0.795 [0.772, 0.813] & 0.754 [0.734, 0.771] & 0.674 [0.652, 0.692] & -0.803 [-0.823, -0.778] & -0.442 [-0.485, -0.397] \\
        QNA entropy & 0.823 [0.803, 0.839] & 0.801 [0.782, 0.816] & 0.737 [0.716, 0.754] & -0.833 [-0.850, -0.812] & -0.489 [-0.527, -0.451] \\
\hline
\textbf{Panel B: Standardized deviations} & \textbf{Rolling-z cov. entropy} & \textbf{Rolling-z eff. rank} & \textbf{Rolling-z mean abs. corr.} & \textbf{Rolling-z realized vol.} & \textbf{} \\
\hline
        QEWS (ERI) & 0.445 [0.397, 0.490] & 0.447 [0.399, 0.491] & -0.422 [-0.469, -0.369] & -0.130 [-0.189, -0.070] &  \\
\hline
\end{tabular}%
}
\end{table}

The more informative comparison concerns \emph{deviations} rather than levels.
The second panel of Table~\ref{tab:benchmark_summary} shows that QEWS, defined as
a rolling standardized deviation of ERI, is only moderately related to rolling-z
covariance spectral entropy ($0.445$) and rolling-z effective rank ($0.447$), and
only weakly related to rolling-z realized volatility ($-0.130$). This matters
for interpretation: level-based QNA measures inherit much of the same regime
information as classical spectra, whereas QEWS is not a relabeling of a
volatility proxy or of rolling covariance entropy.

\begin{figure}[h!]
\centering
\includegraphics[width=0.95\textwidth]{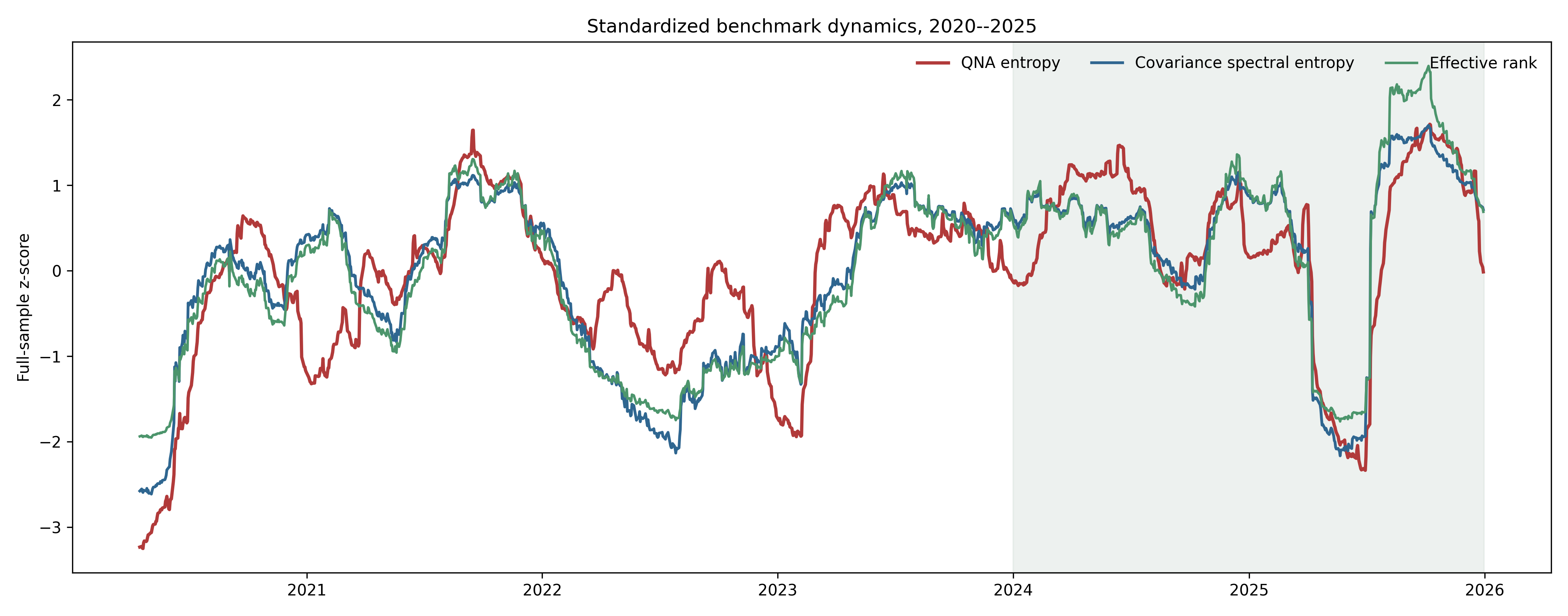}
\caption{Standardized benchmark dynamics in the 2020--2025 main sample. The shaded region marks the 2024--2025 visual spotlight used for the non-crisis zoom-in, while the full series retains the pandemic and 2022 tightening regimes. QNA entropy moves closely with classical spectral benchmarks at the regime level, but not one-for-one.}
\label{fig:benchmark_dynamics}
\label{fig:entropy}
\end{figure}

Figure~\ref{fig:benchmark_dynamics} plots full-sample z-scores of QNA entropy,
covariance spectral entropy, and effective rank over 2020--2025. The broad
co-movement around the 2020 pandemic aftermath, the 2022 tightening cycle, the
2023 banking-stress episode, and the 2024--2025 tariff spotlight is evident. At the
same time, the QNA path is not identical to the classical benchmarks, especially
around 2021 and again in late 2023--2025. These divergences are precisely where
QNA adds value: they arise when dependence is reorganized jointly across return,
volatility, and liquidity channels rather than through contemporaneous return
covariance alone. The shaded 2024--2025 segment is therefore not meant to
exclude 2020 or 2022; it isolates the cleanest non-crisis region in which the
incremental information content of the multi-feature operator can be seen
without the pandemic shock overwhelming the scale of the plot.

\subsection{Sensitivity to Feature Construction and Rolling Windows}

The second result is that the distinct content of QNA depends materially on the
feature space used to construct the operator. This result identifies the domain
in which QNA becomes empirically distinct rather than exposing a nuisance flaw.
Table~\ref{tab:sensitivity_summary}
reports sensitivity checks across three feature specifications and three rolling
windows.

\begin{table}[h!]
\centering
\small
\caption{Sensitivity of QNA diagnostics to feature construction and rolling-window length. The baseline window is 60 trading days; Panel B reports the 40/60/90-day comparison used to preserve stable multi-feature coverage at the short end.}
\label{tab:sensitivity_summary}
\resizebox{\textwidth}{!}{%
\begin{tabular}{lcccccc}
\hline
\textbf{Panel A: Feature specification} & \textbf{Mean dim.} & $\mathbf{E[S_{\mathrm{QNA}}]}$ & $\mathbf{SD[S_{\mathrm{QNA}}]}$ & $\mathbf{Corr(S_{\mathrm{QNA}}, S_{\mathrm{cov}})}$ & $\mathbf{Corr(S_{\mathrm{QNA}}, r_{\mathrm{eff}})}$ & $\mathbf{Corr(QEWS, zS_{\mathrm{cov}})}$ \\
\hline
        Returns only & 60.0 & 2.850 & 0.416 & 0.966 & 0.940 & 0.900 \\
        Returns + volatility & 118.9 & 2.725 & 0.440 & 0.897 & 0.858 & 0.648 \\
        Baseline full & 237.8 & 3.211 & 0.301 & 0.823 & 0.801 & 0.445 \\
\hline
\textbf{Panel B: Rolling window} & \textbf{Mean dim.} & $\mathbf{E[S_{\mathrm{QNA}}]}$ & $\mathbf{SD[S_{\mathrm{QNA}}]}$ & $\mathbf{Corr(S_{\mathrm{QNA}}, S_{\mathrm{cov}})}$ & $\mathbf{Corr(S_{\mathrm{QNA}}, r_{\mathrm{eff}})}$ & $\mathbf{Corr(QEWS, zS_{\mathrm{cov}})}$ \\
\hline
        40-day & 158.4 & 3.101 & 0.319 & 0.782 & 0.746 & 0.390 \\
        60-day & 237.8 & 3.211 & 0.301 & 0.823 & 0.801 & 0.445 \\
        90-day & 357.5 & 3.295 & 0.297 & 0.858 & 0.849 & 0.534 \\
\hline
\end{tabular}%
}
\end{table}

The feature comparison is especially revealing. When QNA is constructed from
\emph{returns only}, QNA entropy is almost isomorphic to covariance spectral
entropy (correlation $0.966$) and QEWS tracks the rolling-z covariance benchmark
very closely (correlation $0.900$). Adding volatility and then liquidity/activity
channels weakens those links monotonically, with the baseline multi-feature
specification falling to $0.823$ for the level comparison and $0.445$ for the
QEWS comparison. This is direct empirical evidence that QNA coincides with
classical spectral methods in the returns-only limit and departs from that limit
as additional feature channels are introduced. The richer multi-feature operator
is therefore not an arbitrary embellishment; it is the mechanism through which
QNA captures dependence geometry beyond return covariance alone.

The rolling-window results are more stable. Across $40$-, $60$-, and $90$-day
windows, the mean level of QNA entropy shifts mechanically with operator
dimension, but the qualitative benchmark relationship remains intact:
correlations with covariance spectral entropy stay in the $0.78$--$0.86$ range
and correlations between QEWS and rolling-z covariance entropy remain moderate
($0.39$--$0.53$). We therefore interpret QNA as \emph{design dependent but not
fragile}: the exact level of the diagnostics depends on feature construction and
window length, while the broader benchmark structure is stable. The 40-day
window is intentionally retained as the shortest robustness horizon because it
is close to short-horizon event-study practice while avoiding the additional
coverage instability induced by a fully multi-feature 30-day operator; it is
noisier than the 60-day baseline, but it does not reverse the central empirical
ordering.

\subsection{Event-Window Evidence and the Tariff Cluster Case}

The original manuscript used the February~2025 U.S.\ tariff announcement as a
single illustrative case. The revised event design instead centers on dated,
broad macro and policy repricing windows: Fed liftoff (March~16, 2022), the
Jackson Hole hawkish speech (August~26, 2022), the February~18, 2025 tariff
announcement, the April~2, 2025 reciprocal tariff escalation, and two calm
placebo dates. Because QEWS requires a 60-day rolling
normalization window, the March~2020 pandemic announcement remains visible in
Figure~\ref{fig:benchmark_dynamics} but is not used in the formal event table.
We also screened more localized bank-stress windows, such as the March~2023 SVB
failure; those responses are materially weaker and are therefore treated as
boundary cases rather than as headline validation.

\begin{table}[h!]
\centering
\small
\caption{Event-window evidence for QEWS (ERI) across selected macro/policy and placebo windows. Pre- and post-event means use a symmetric 20-trading-day window. The reported difference is pre minus post, with bootstrap confidence intervals and a permutation $p$-value.}
\label{tab:event_summary}
\resizebox{\textwidth}{!}{%
\begin{tabular}{lcccccc}
\hline
\textbf{Event} & \textbf{Date} & \textbf{Type} & \textbf{Pre QEWS} & \textbf{Post QEWS} & \textbf{Pre - Post} & \textbf{95\% CI / Perm. $p$} \\
\hline
        Fed rate liftoff & 2022-03-16 & macro & -1.566 & -0.182 & -1.385 & [-1.855, -0.898]; p=0.000 \\
        Jackson Hole speech & 2022-08-26 & macro & 0.664 & 2.175 & -1.511 & [-1.956, -1.032]; p=0.000 \\
        US tariff announcement & 2025-02-18 & policy & -0.704 & -0.113 & -0.591 & [-0.894, -0.256]; p=0.002 \\
        Reciprocal tariff escalation & 2025-04-02 & policy & 0.245 & -2.643 & 2.889 & [2.009, 3.773]; p=0.000 \\
        Placebo window A & 2023-06-06 & placebo & 1.077 & 0.918 & 0.159 & [-0.189, 0.518]; p=0.395 \\
        Placebo window B & 2024-06-07 & placebo & 0.603 & 0.680 & -0.077 & [-0.940, 0.708]; p=0.870 \\
\hline
\end{tabular}%
}
\end{table}

\begin{figure}[H]
\centering
\includegraphics[width=0.95\textwidth]{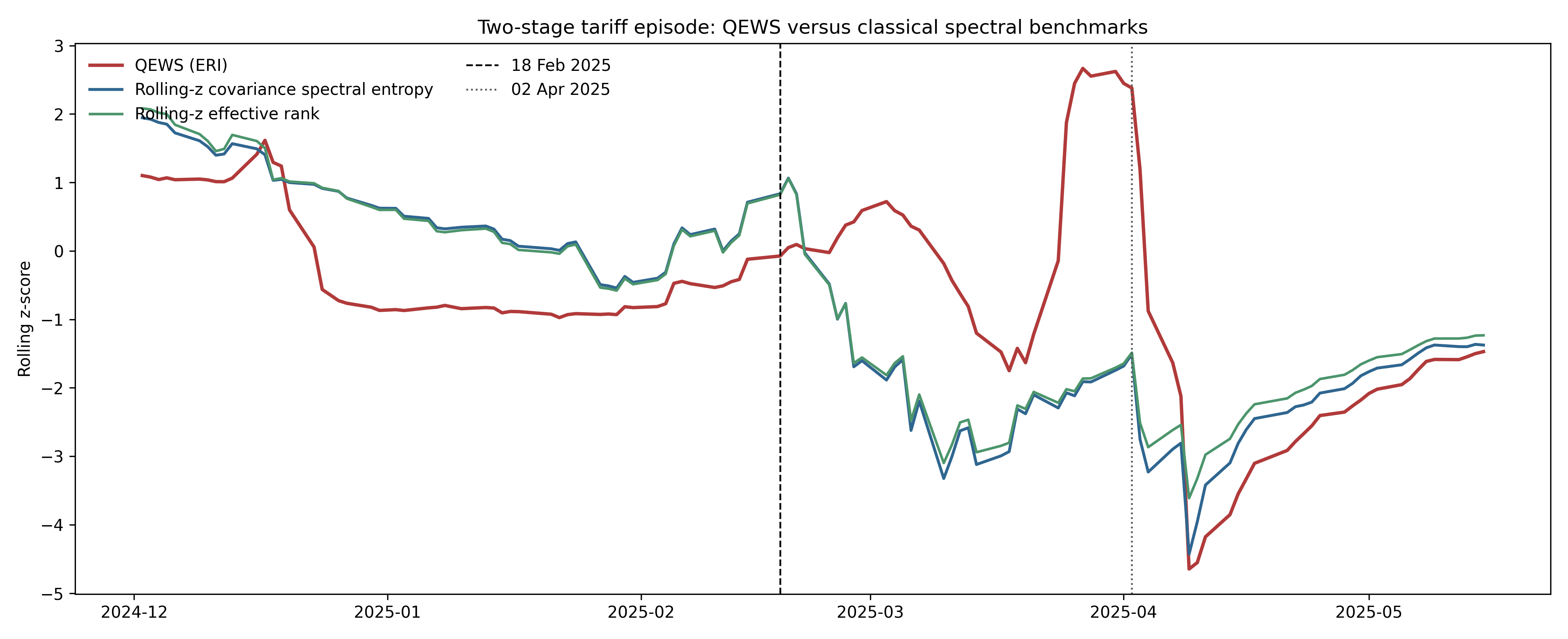}
\caption{Two-stage tariff episode, December~2024 to May~2025. The dashed and dotted lines mark the February~18 initial tariff announcement and the April~2 reciprocal tariff escalation. QEWS (ERI) is compared with rolling-z covariance spectral entropy and rolling-z effective rank.}
\label{fig:tariff_focus}
\label{fig:qews}
\end{figure}

Three points follow from Table~\ref{tab:event_summary}. First, the strongest
headline responses occur around market-wide macro/policy repricing rather than
around localized institution-specific stress: QEWS shifts are statistically
significant around Fed liftoff ($-1.385$), Jackson Hole ($-1.511$), the initial
tariff announcement ($-0.591$), and especially the April tariff escalation
($2.889$). Second, the placebo dates do not generate statistically significant
QEWS shifts, which helps discipline the interpretation of the positive cases.
Third, the sign of the QEWS change is not uniform across events, reinforcing the
earlier point that QEWS should be read as a signed measure of abnormal
structural deviation rather than as a monotone ``stress goes up'' index.

Figure~\ref{fig:tariff_focus} retains the tariff episode because it occurs in a
comparatively non-crisis environment and therefore offers a cleaner view of
structural reconfiguration than the pandemic period. The revised figure is now
organized as a two-stage tariff cluster and compares QEWS against rolling-z
covariance spectral entropy and rolling-z effective rank, which are the most
direct classical comparators. The February~18 announcement generates movement in
both QEWS and the classical spectral benchmarks, consistent with a common
first-stage repricing. The April~2 escalation is different: QEWS shifts sharply while
the rolling-z spectral comparators move only modestly. That divergence is where
QNA is most informative. It is the clearest empirical validation in the paper:
the multi-feature operator registers renewed dependence reconfiguration rather
than a simple continuation of spectral concentration, which is precisely the
incremental value claimed for the QNA representation.

Taken together, the empirical evidence supports a narrower and more defensible
claim than in the original submission. QNA is strongest as an operator-based
framework that unifies entropy, structural mixing, and event-aligned deviations
within one representation. It is not a blanket replacement for covariance
spectral methods; rather, it becomes most distinct when the operator is built on
multi-feature rolling trajectories and then used as a structural diagnostic.

\section{Discussion}

Section~5 supports three discussion points. First, QNA is best interpreted as an
\emph{operator-based representation change}, not as an empirical repudiation of
rolling PCA or covariance-spectrum analysis. The strong level correlations with
covariance spectral entropy and effective rank confirm that QNA shares their
effective-dimensionality content. The contribution of QNA is instead to provide
a single operator object in which entropy, purity-based mixing, standardized
deviations, and partition-based extensions can be analyzed jointly across
rolling multi-feature trajectories.

Second, the empirical distinctiveness of QNA is concentrated in the
\emph{multi-feature} specification. The returns-only case behaves almost like a
normalized covariance-spectrum diagnostic, which is exactly the identification
boundary the paper needs. Once volatility and liquidity/activity paths are added
to the rolling block, the QNA operator departs from the returns-only spectral
limit and begins to encode a broader dependence geometry. This is the clearest
answer to the reviewer concern that QNA might simply rename PCA: it converges
toward classical spectral methods in the returns-only limit and becomes
empirically distinct when heterogeneous channels are aggregated into a common
operator state.

Third, the event evidence clarifies \emph{where} the method is most informative.
QEWS is strongest in the revised catalog around broad macro/policy repricing and
most clearly during the April~2025 tariff escalation, where QNA identifies a
new structural deviation while rolling-z spectral benchmarks move only modestly.
This is why the manuscript visually emphasizes 2024--2025 while retaining the
full 2020--2025 sample: the pandemic period is essential for understanding
extreme dependence concentration, but the 2025 tariff cluster provides a cleaner
setting in which incremental structural reconfiguration can be distinguished from
broad volatility shocks. QEWS should therefore be interpreted as a signed
deviation from a recent dependence baseline rather than as a monotonic measure
of systemic stress.

These points carry two practical implications for systemic-risk monitoring.
First, operator-based diagnostics may help track the erosion of diversification
capacity when fewer common modes dominate the cross-section. Second, because the
framework works with standardized rolling feature blocks, it provides a common
language in which dependence concentration, structural mixing, and abnormal
reconfiguration can be discussed jointly instead of through disconnected summary
statistics. Mutual information remains a natural extension of this framework,
but the empirical section intentionally focuses on entropy, ERI, and QEWS in
order to isolate the main contribution rather than dilute it across partition
design choices.

The limitations are equally clear. The current implementation uses a stable
NASDAQ--100 panel anchored to current membership and therefore inherits
survivorship bias. The diagnostics depend on feature design and window length by
construction. Eigenvalue-based quantities can also be sensitive to finite-sample
noise when the operator dimension grows relative to the window length. Finally,
QNA is descriptive: it summarizes dependence compactly, but it does not identify
causal channels or replace economic modeling of the shocks that move the system.
Future work should therefore focus on historical constituent reconstruction or
alternative universes, regularized inference for the density operator, and more
systematic partition design for mutual-information decomposition. Those
extensions would help determine more precisely when QNA delivers distinct
empirical information and when it reduces to classical spectral summaries.

\section{Conclusion}

This paper proposes the \emph{Quantum Network of Assets} (QNA) as an
operator-based framework for analysing cross-asset dependence through normalized
rolling feature blocks. The framework yields two structural diagnostics -- the
Entanglement Risk Index (ERI) and the Quantum Early-Warning Signal (QEWS) -- and
uses them to describe effective dimensionality, structural mixing, and abnormal
departures from recent dependence configurations.

The revised empirical results support a sharper interpretation of the method.
QNA entropy remains strongly related to covariance spectral entropy and effective
rank, so QNA should be viewed as a complementary spectral representation rather
than as a rejection of PCA-style tools. The feature-sensitivity exercises show
more specifically that returns-only QNA lies close to the classical spectral
limit, whereas the multi-feature operator is where QNA becomes empirically
distinct. The broader event-window design further shows that QEWS is most
informative around broad macro/policy repricing, with the April~2025 tariff
escalation providing the clearest case of structural deviation beyond what the
rolling spectral benchmarks alone capture.

The main contribution of QNA is therefore representational but not merely
terminological. It places entropy, purity-based mixing, and event-aligned
structural deviations inside a single operator framework, and it shows most
clearly where that representation becomes valuable: when dependence is built
from rolling multi-feature trajectories rather than from return covariance
alone. Within its boundaries -- design dependence, finite-sample sensitivity,
and survivorship bias in the stable-panel implementation -- QNA provides a
compact language for describing the time-varying geometry of market dependence
and complements existing covariance-spectrum diagnostics in financial risk
analysis.

\section*{Acknowledgments}
The authors gratefully acknowledge the support of the UCL Institute of Finance
and Technology (IFT). The authors also thank Prof. Thomas Schroeder, Industry
Professor at IFT, for insightful discussions and exchanges that significantly
contributed to a deeper understanding of the underlying concepts explored in
this work. Any remaining errors are the authors’ own responsibility.


\end{document}